ARTICLE

# Site-specific reactivity of ethylene at distorted dangling bond configurations on Si(001)

Josua Pecher,[a] Gerson Mette,[b] Michael Dürr*[b,c] and Ralf Tonner*[a]

**Abstract:** We report differences in adsorption and reaction energetics for ethylene on Si(001) with respect to different dangling bond configurations induced by hydrogen precoverage as obtained via density functional theory calculations. This can help to understand the influence of surface defects and precoverage on the reactivity of organic molecules on semiconductor surfaces in general. Our results show that the reactivity on surface dimers fully enclosed by hydrogen covered atoms is essentially unchanged compared to the clean surface. This is confirmed by our scanning tunnelling microscopy measurements. On the contrary, adsorption sites with partially covered surface dimers show a drastic increase in reactivity. This is due to a lowering of the reaction barrier by more than fifty percent compared to the clean surface, which is in line with previous experiments. Adsorption on dimers enclosed by molecule (ethylene) covered surface atoms is reported to have a highly decreased reactivity, a result of destabilization of the intermediate state due to steric repulsion, as quantified with the periodic energy decomposition analysis (pEDA). Furthermore, an approach for the calculation of Gibbs energies of adsorption based on statistical thermodynamics considerations is applied to the system. The results show that the loss in molecular entropy leads to a significant destabilization of adsorption states.

## 1. Introduction

The chemistry of organic molecules on surfaces has been a major topic in material science research for several years.[1] With respect to the development of new materials and electronic devices, the organic functionalization of group 14 (aka. group IV) semiconductor surfaces is of great interest.[2] Within this group, silicon is most interesting due to its wide-spread use and applications in the electronics industry and a great number of organic molecules reacting on silicon surfaces have already been investigated.[3] The Si(001) surface is especially suited for this since it is known to form dimers[4] that show a pronounced chemical reactivity.[5]

Ethylene on Si(001) has already been studied extensively in experiment[6] and theory.[7] The reaction mechanism is uniformly agreed upon: Scheme 1 shows the weakly bound π complex, henceforth called precursor, which is the predominant structure at low temperatures and a short-lived intermediate at higher temperatures, and the two possible covalently bound structures to which the precursor can convert: *on-top* on a single dimer and *bridge* between two adjacent dimers. Although both products are accessible in the course of the reaction, calculations[7h,7s] have shown the energy barrier from the precursor to the *bridge* structure to be significantly higher (11-16 kJ mol$^{-1}$) than to the *on-top* structure (2-7 kJ mol$^{-1}$). In previous scanning tunneling microscopy (STM) measurements by some of the authors the *bridge* structure has been observed experimentally in addition to the predominant *on-top* structure.[6r] While the existence of the *bridge* structure was then also confirmed by spectroscopic measurements,[6s] former STM studies did not observe the *bridge* structure.[6c,6n,6q] For the low temperature study[6q] this can be attributed to the aforementioned significant difference in barriers. Furthermore, our experiments also showed an increased reactivity towards *bridge* states in surface areas partially covered with hydrogen.[6r] This result is puzzling and was the driving force for the present study.

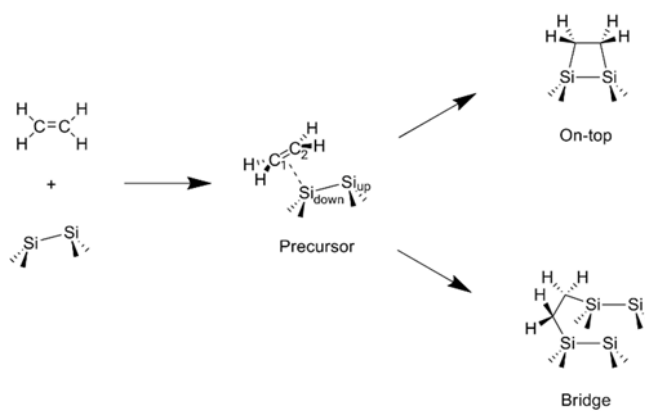

**Scheme 1.** Reaction pathway of ethylene on the Si(001) surface: Adsorption into the π complex (precursor) on a single surface dimer at low temperatures, conversion into either the one-dimer (*on-top*) or two-dimer (*bridge*) covalently bound states at elevated temperatures.

All theoretical studies up to now are restricted to the reactivity of ethylene with the clean surface. We now aim at analyzing the influence of hydrogen precoverage to explain the increased reactivity observed experimentally. This is important for understanding how defects, impurities and already adsorbed atoms or molecules affect the dynamics of the adsorption process.

[a] Josua Pecher, Dr. Ralf Tonner
Faculty of Chemistry and Material Sciences Centre
Philipps-Universität Marburg
Hans-Meerwein-Str. 4, D-35032 Marburg, Germany
E-mail: tonner@chemie.uni-marburg.de

[b] Dr. Gerson Mette, Prof. Dr. Michael Dürr
Faculty of Physics and Material Sciences Centre
Philipps-Universität Marburg
Renthof 5, D-35032 Marburg, Germany

[c] Prof. Dr. Michael Dürr
Institute of Applied Physics
Justus Liebig University Giessen
Heinrich-Buff-Ring 16, D-35392 Giessen, Germany
E-mail: michael.duerr@ap.physik.uni-giessen.de

Supporting information for this article is given via a link at the end of the document.

In doing so, density functional theory (DFT) with semiempirical dispersion correction (DFT-D3) is being used. While the surface could be modeled by either a cluster or a periodic slab, we chose the latter option as it describes the electronic situation of the surface more accurately. The dispersion treatment is needed to correct for the known failure of most density functionals in describing these interactions.[8]

Furthermore, new experimental results for the reactivity of ethylene at isolated dimers for high hydrogen and ethylene precoverage are presented and discussed. Using bonding analysis methods, the observed differences between the different surfaces can be easily explained.

## 2. Methods

### 2.1. Computational Details

All calculations have been performed with periodic DFT as implemented in the *Vienna Ab-initio Simulation Package* (VASP),[9] version 5.3.5, using the projector augmented-wave (PAW) formalism[10] and the exchange-correlation functional by Perdew, Burke and Enzerhof (PBE).[11] To account for dispersive interactions, the semi-empirical DFT-D3 correction by Grimme and co-workers[12] was used. The basis set consisted of plane waves up to a cutoff corresponding to a kinetic energy of 400 eV, while electronic $k$ space has been sampled using a gamma-centered Monkhorst-Pack grid, where a $\Gamma(241)$ grid was used for the 4×2 sized cells and a $\Gamma(221)$ grid for the 4×4 cells. Convergence criteria for the self-consistent field calculations and structural optimizations have been chosen as $10^{-5}$ eV and $10^{-2}$ eV/Å, respectively, while for structures to be used in frequency calculations, those values were chosen as $10^{-8}$ eV and $10^{-3}$ eV/Å. The Conjugate Gradient optimization algorithm[13] was used for structural optimizations. All obtained geometries are given in the Supplementary Information.

The silicon surface was modeled in a slab approach with a thickness of six layers, where the atoms at the bottom two layers were frozen to their bulk positions and saturated with hydrogen atoms, pointing in the direction of the next bulk layer atoms, at a distance of $d$(Si-H) = 1.480 Å, the experimental equilibrium Si-H bond length in Silane.[14] The unit cell of the $c$(4×2) surface reconstruction with buckled dimers (Scheme 2, left) was used for the *on-top* reactions, while a (1×2) supercell was used for the *bridge* reactions. This is equal to a coverage of θ = 0.25 (*on-top*) and θ = 0.125 (*bridge*) molecules per surface dimer, respectively. To minimize interaction with periodically repeated images of the slab in c direction, a vacuum layer of at least 10 Å was ensured. The $a$ and $b$ cell constants were set to 15.324 and 7.662 Å, respectively, derived from our optimized bulk lattice parameter of $a$ = 5.418 Å. Hydrogen-precovered surfaces were obtained by adding hydrogen atoms to the slab structures and subsequent structural optimization. The **Si-ID** surface (Scheme 2, center) was chosen to resemble an isolated dimer (ID) in an environment of hydrogen-saturated dimers, representing high H-precoverage. The **Si-H4** surface (Scheme 2, right; see Dürr *et al.*[15] for nomenclature) was constructed according to the observed features from our previous experiments.[6r]

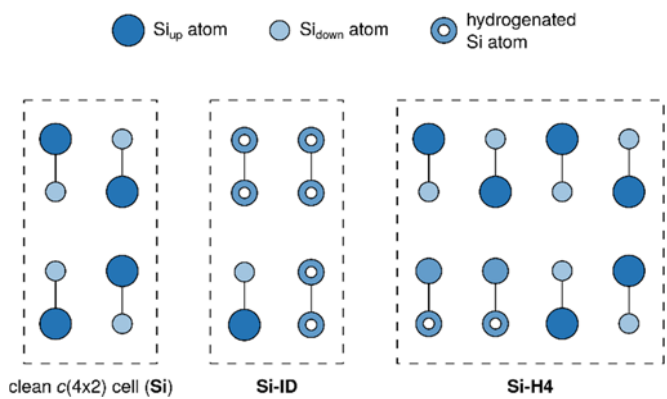

**Scheme 2.** Unit cell of the clean Si(001) surface and the two hydrogen precovered H/Si(001) surfaces **Si-ID** and **Si-H4** investigated. Cell boundaries shown in dashed lines.

Adsorption energies $E_{ads}$ are given as the difference between the energy of the optimized structure for the total system $E_{tot}$ and the relaxed and isolated molecule and slab energies $E_{mol}$ and $E_{slab}$:

$$E_{ads} = E_{tot} - E_{mol} - E_{slab} \quad (1)$$

Reaction pathways and their corresponding energy profiles have been calculated using the Nudged Elastic Band method (NEB) with the climbing image modification.[16] All NEB calculations have been done using the limited-memory Broyden-Fletcher-Goldfarb-Shanno optimization algorithm,[17] as this has been shown to be the most efficient algorithm in these kind of calculations.[18] Interpolation in the energy profiles was done using a cubic spline based on the forces along the reaction coordinate.

Bonding analysis was carried out with the PBE-D3 method, a TZ2P basis set and sampling electronic k-space at the Γ-point using ADF-BAND 2016.[28] A description of the periodic Energy Decomposition Analysis method (pEDA) is found in Ref. [29].

### 2.2. Theoretical Treatment of Finite Temperature and Pressure Effects

Thermodynamic correction terms for enthalpy ($H_{corr}$) and entropy ($S_{corr}$) were applied to the systems using the well-known equations for the harmonic oscillator, rigid rotor and ideal gas from statistical thermodynamics. This is necessary because entropy changes play a huge role in adsorption processes of molecules on surfaces[19] and are not described by single-point DFT calculations which only give the electronic energy $E_{el}$. Accordingly, the Gibbs energy $G$ was calculated for all structures using the following equations:[20]

In these equations, $T$ denotes the temperature, $v_i$ the computed vibrational frequencies, $V$ the occupied volume of a single molecule, $m$ the molecular mass, $\sigma$ the symmetry number, and $I_1$, $I_2$ and $I_3$ the molecule's moments of inertia. In case of constant pressure conditions instead of constant volume, $V$ can be substituted by $kT/p$ assuming ideal gas behavior.

Harmonic frequencies were calculated from the Hessian matrix, which was constructed using a finite difference approach with displacements of 0.01 Å from the equilibrium structure. For the isolated molecule, the optimized structure and vibrational frequencies were used from a calculation in a cell sized equally to the respective reaction (4×2 or 4×4). This was done to avoid artificial errors due to the change in basis set size and effects of the periodic boundary conditions.

$$G = E_{\text{el}} + H_{\text{corr}} + T S \qquad (2)$$
$$H_{\text{corr}} = H_{\text{vib}} + H_{\text{rot}} + H_{\text{trans}} \qquad (3)$$
$$S = S_{\text{vib}} + S_{\text{rot}} + S_{\text{trans}} \qquad (4)$$
$$H_{\text{vib}} = R \sum_i^{3N} \left( \frac{h\nu_i}{k} \left( \frac{1}{2} + \frac{1}{e^{h\nu_i/kT} - 1} \right) \right) \qquad (5)$$
$$H_{\text{rot}} = \frac{3}{2} RT \qquad H_{\text{trans}} = \frac{5}{2} RT \qquad (6)$$
$$S_{\text{vib}} = R \sum_i^{3N} \left( \frac{h\nu_i}{kT} \frac{1}{e^{h\nu_i/kT} - 1} - \ln(1 - e^{-h\nu_i/kT}) \right) \qquad (7)$$
$$S_{\text{rot}} = R \left( \frac{3}{2} + \ln \left( \frac{\sqrt{\pi}}{\sigma} \left( \frac{8\pi^2 kT}{h^2} \right)^{\frac{3}{2}} \sqrt{I_1 I_2 I_3} \right) \right) \qquad (8)$$
$$S_{\text{trans}} = R \left( \frac{5}{2} + \ln \left( V \left( \frac{2\pi m kT}{h^2} \right)^{\frac{3}{2}} \right) \right) \qquad (9)$$

The temperature was chosen to be $T = 300$ K to be comparable to experiments, which were mostly done at room temperature. Nonetheless, we show electronic energies side-by-side with Gibbs energies to enable discussion of the temperature dependence of the energy values. Since experiments are usually done at ultra-high vacuum conditions, but the local pressure can vary by many orders of magnitude, multiple aspects have to be considered in the choice of a pressure value: State-of-the-art Gibbs energy calculations are done using a Potential of Mean Field (PMF) within Molecular Dynamics simulations.[21] In order to get reliable values and good sampling of all degrees of freedom, long simulation times are needed, which can only be achieved by using empirical force fields. Since there is currently no reliable force field available for this kind of surface, this approach cannot be pursued in our case. Assuming the harmonic oscillator and rigid rotor approximations appropriately describe the behavior that can be found in PMF calculations, translation will be modeled in the following way: PMF calculations define the Gibbs energy as a function of the molecule-surface distance with this property being constrained during simulation. At large distances, translation in the other two directions should ideally behave like a one-molecule 2D ideal gas confined to an area $A$ spanned by the cell vectors $a$ and $b$. Consequently, we decided that an appropriate approximation of the translation would be to treat the molecule as a 3D ideal gas confined to the available volume $V_{\text{avail}}$ of the cell with $V_{\text{avail}} = V_{\text{cell}} - V_{\text{slab}}$. While this is dependent on the choice of the cell and slab size, the property $V$ is included in a logarithmic expression, so only the order of magnitude should be relevant. In our case, this is also very close to the ideal gas value at 1 bar and 300 K, at which standard condition entropies are usually given in surface adsorption experiments.[19b] Neglecting these small errors, we consistently chose $p = 1$ bar to be coherent with experiments. In flat regions of the potential energy surface, hindered rotation and translation can become relevant and entropy terms might not be described accurately enough anymore in the harmonic oscillator approximation. Therefore, we checked our structures for these motions. Only one hindered rotation and no hindered translation were found to have an energy barrier smaller than the barriers leading to the covalently bound structures. The thermodynamic correction terms of this motion were calculated once treating it as a harmonic oscillator and once as a free rotor, both limiting cases. The value of the real system should lie between these values. The results (Supplementary Information, Table S2) showed the overall change to be only 2 kJ mol$^{-1}$, which is lower than other methodological errors, e.g. general errors of DFT, so we decided to keep the harmonic oscillator approximation. The same approach was used in a detailed analysis of the precursor structure.[30]

### 2.3. Experimental Methods

The experiments were performed using a commercial OMICRON VT-STM in an ultrahigh vacuum chamber with a base pressure below 1×10$^{-10}$ mbar. The n-doped Si(001) samples oriented within 0.25° along the (001) direction were prepared by degassing the sample at 700 K and repeatedly flashing to temperatures above 1450 K by means of direct current heating. Slow cooling down to room temperature with rates of about 1 K s$^{-1}$ then results in a clean and well-ordered Si(001) 2×1 reconstruction with a minimum of defects.[22] Hydrogen precovered surfaces were prepared by dosing highly purified H$_2$ gas (99.9999% purity) via a gas inlet system equipped with a liquid nitrogen trap to freeze out residual impurities. Molecular hydrogen was dissociated at a hot tungsten filament (approx. 2000 K) which was positioned 5 cm from the sample. Typical exposures were 2.5×10$^{-6}$ mbar H$_2$ gas for 660 s. Ethylene with 99.95% purity was dosed via a second gas inlet system with exposures of 0.15 L to 0.65 L of C$_2$H$_4$ gas (1 L = 1.33 ×10$^{-6}$ mbar s). During the ethylene exposure the STM tip was withdrawn from the sample. All experiments were performed at room temperature, ion gauge readings were corrected for relative ionization probabilities.

## 3. Results and Discussion

### 3.1. Computational Results

The optimization of the isolated ethylene molecule resulted in structural parameters of $r_{\text{CC}} = 1.333$ Å, $r_{\text{CH}} = 1.092$ Å, $α_{\text{HCH}} = 116.6°$ and $α_{\text{HCH}} = 121.7°$, which are in excellent agreement with experimental data[23] (1.339/1.086 Å, 117.2/121.2°) and molecular DFT calculations at the similar PBE/TZVP level[24] (1.334/1.092 Å, 116.5/121.8°). To validate the quality of our frequency calculations, comparison with literature showed our values for the isolated molecule to be in good agreement with experiment[25] and PBE/TZVP calculations[24] (Supplementary Information, Table S1), which give zero point vibrational energies of 10785 (experiment) and 10888 cm$^{-1}$ (theory), while our calculations yield values of 10893 cm$^{-1}$ (4×2 cell) and 10889 cm$^{-1}$ (4×4 cell).

Table 1 shows the individual adsorption energies $E_{\text{ads}}$ and Gibbs energies $G_{\text{ads}}$ for the precursor, transition state and final state in all four reactions considered. Figure 1 shows an illustration of the different energy terms discussed for an exemplary potential energy curve of adsorption, which depicts the actual energy profile for the adsorption into the *bridge* structure on the clean **Si** surface, as calculated with the NEB method. The large differences between electronic and Gibbs energies in Table 1 (ca. +45-65 kJ mol$^{-1}$), can mainly be attributed to the loss of translational and rotational entropy of the molecule due to adsorption on the surface ($\Delta TS_{trans}$ = +45, $\Delta TS_{rot}$ = +20 kJ mol$^{-1}$, see also Supplementary Information, Table S3), which are only partially compensated by the increase in vibrational entropy ($\Delta TS_{vib}$ = −19 kJ mol$^{-1}$). This example shows that using electronic energies can lead to a major overestimation of the adsorption and desorption energies for many surface adsorption reactions.

**Table 1.** Electronic and Gibbs energies of bonding of the precursor ($E_{prec}$), transition state ($E_{TS}$) and final state ($E_{final}$) in the four reactions considered. The energies relative to the respective precursor ($E^{rel}_{TS}$, $E^{rel}_{final}$) are given as well. See also Figures 2 and 3 for corresponding structures.[a]

|       |       | $E_{prec}$ | $E_{TS}$ | $E_{final}$ | $E^{rel}_{TS}$ | $E^{rel}_{final}$ |
|-------|-------|-----|-----|------|----|------|
|       |       | *on-top* | | | | |
| $E_{ads}$ | **Si**    | −74 | −66 | −201 | 9  | −127 |
|       | **Si-ID** | −62 | −54 | −213 | 8  | −151 |
| $G_{ads}$ | **Si**    | −22 | −12 | −141 | 10 | −119 |
|       | **Si-ID** | −13 | −3  | −156 | 10 | −143 |
|       |       | *bridge* | | | | |
| $E_{ads}$ | **Si**    | −74 | −57 | −184 | 17 | −110 |
|       | **Si-H4** | −79 | −71 | −238 | 8  | −159 |
| $G_{ads}$ | **Si**    | −22 | −5  | −117 | 17 | −95  |
|       | **Si-H4** | −26 | −19 | −175 | 7  | −148 |

[a] All values in kJ mol$^{-1}$, calculated using PBE-D3/PAW. Gibbs energies at $T$ = 300 K, $p$ = 1 bar.

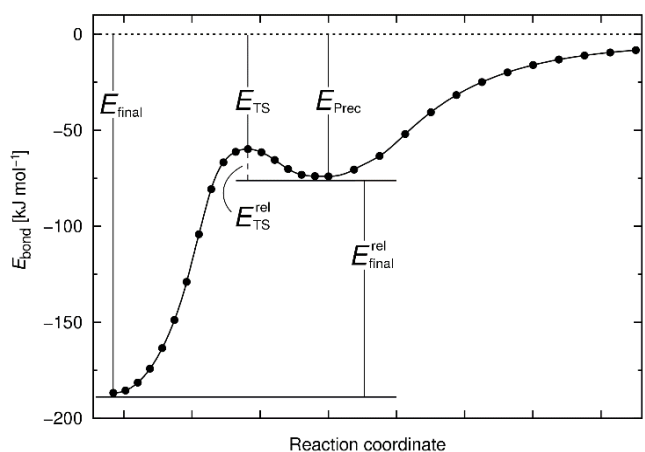

**Figure 1.** Potential energy curve of the adsorption of ethylene into the *bridge* structure on clean Si(001), calculated with the NEB method, introducing the energy terms discussed.

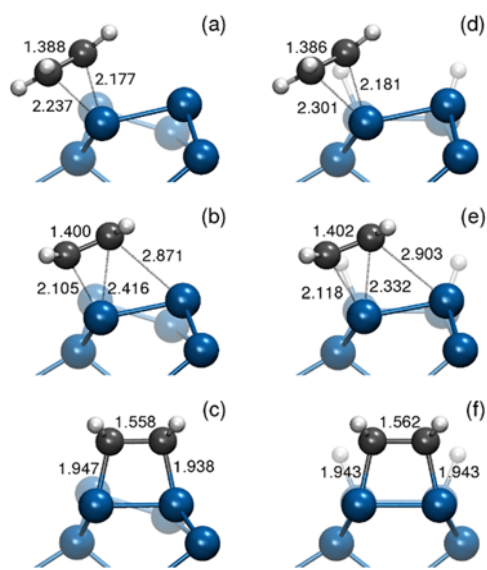

**Figure 2.** Pathway into the *on-top* structure: Precursor (a,d), transition state (b,f) and final state (c,f) structure for the reaction on the Si (a-c) and Si-ID (d-f) surface. Bond lengths in Å. Transition state imaginary frequencies: 138$i$ cm$^{-1}$ (b) and 153$i$ cm$^{-1}$ (e).

The nomenclature of the carbon and silicon atoms involved in the reaction, as it will be used in the discussion of interatomic distances in the following section, is depicted in Scheme 1.

### 3.1.1. The Precursor State

The three different precursor geometries are shown in Figures 2(a) (**Si**), 2(d) (**Si-ID**) and 3(d) (**Si-H4**). There are no noticeable differences in the molecular orientation and in all cases, the C-C bond is not in plane with the Si-Si dimer bond, but significantly rotated (see Figure 4 for a definition of the rotation angle): **Si** and **Si-ID**: 37°; **Si-H4**, 33°. This is in agreement with a previous *ab-initio* Molecular Dynamics study[7w] that showed a maximum of the angular distribution at about 45°. As our calculated energy profile for the hindered rotation on the clean surface shows (Figure 4), the 0° orientation is actually a first-order saddle point ($\nu_{imag}$ = 60$i$ cm$^{-1}$), with a small energy difference of about 2 kJ mol$^{-1}$ to the minimum, while the 90° orientation is a very shallow minimum at about 3 kJ mol$^{-1}$.

A look at the interatomic distances (Figures 2(a,d) and 3(a,d)) shows that in all cases, the C-C bond length is elongated from 1.333 Å in the gas phase to 1.386-1.389 Å in the precursor structures, indicating that this bond is already weakened in this state. While a symmetric coordination of the C-C bond to the Si$_{down}$ might be expected in the first place, the different Si-C bond lengths show that on **Si-H4**, the C$_1$ atom is closer to the coordinated Si$_{down}$ compared to C$_2$ by 0.034 Å, while the **Si** and **Si-ID** precursors have the C$_2$ atom closer by 0.060/0.055 and 0.120 Å, respectively. These values also show that the **Si-ID** precursor is more asymmetrically bound with respect to the carbon atoms, which can be attributed to steric repulsion to the hydrogen atom on the adjacent dimer, missing for the other two surface configurations. Precursor bonding energies (Table 1) show a value of −74 kJ mol$^{-1}$ on the clean surface for both reactions. Given that the only difference in those two calculations is the cell size, this implies that the 4×2 cell is large enough to properly describe the bonding situation in this system.

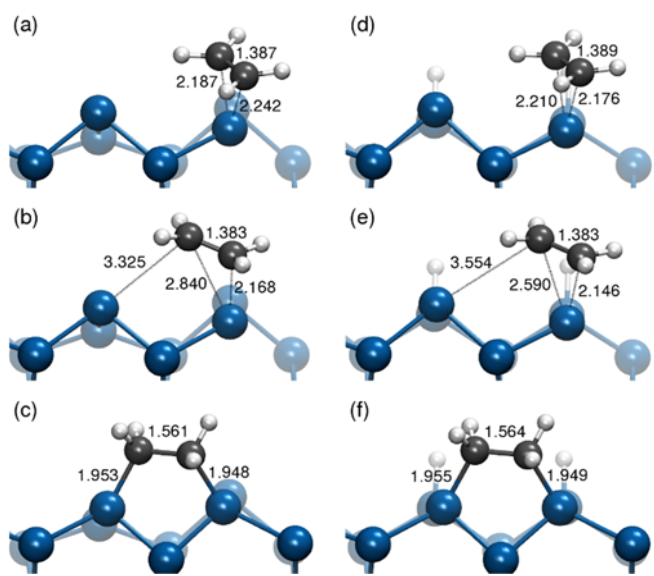

**Figure 3.** Pathway into the *bridge* structure: Optimized precursor (a,d), transition state (b,e) and final state (c,f) structure for the reaction on the **Si** (a-c) and **Si-H4** (d-f) surface. Bond lengths in Å. Transition state imaginary frequencies: 103$i$ cm$^{-1}$ (b) and 125$i$ cm$^{-1}$ (e).

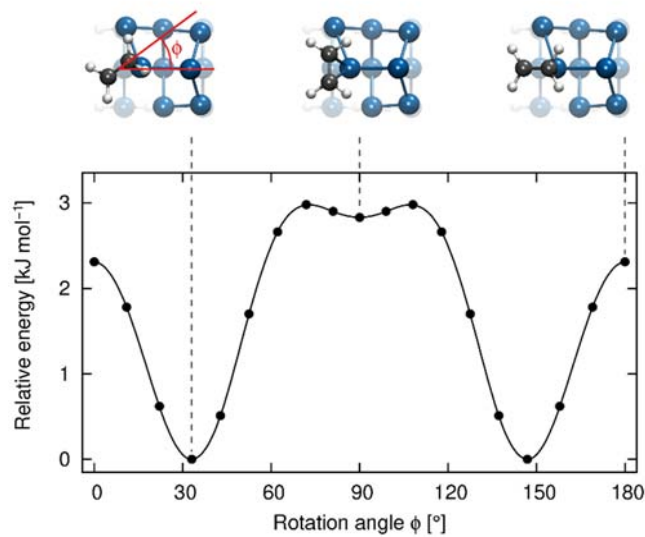

**Figure 4.** Energy profile of rotation for the precursor on the **Si** surface as calculated with the NEB method. Zero point set at the minimum energy orientation. $\phi$ denotes the angle between the $y$ axis and the C-C bond axis projected onto the $xy$ plane.[26]

Precursor bonding energies (Table 1) show a value of −74 kJ mol$^{-1}$ on the clean surface for both reactions. Given that the only difference in those two calculations is the cell size, this implies that the 4×2 cell is large enough to properly describe the bonding situation in this system. The **Si-H4** precursor is more strongly bound by 4-5 kJ mol$^{-1}$, while the **Si-ID** precursor is weaker by 9-12 kJ mol$^{-1}$, a trend that is reflected in the average C-Si bond lengths of 2.207/2.215 (**Si**), 2.241 (**Si-ID**) and 2.193 Å (**Si-H4**). Comparing electronic to Gibbs energies, the aforementioned positive shift of bonding energies is observed and the precursor on the clean surface (+52 kJ mol$^{-1}$) is similar to the ones on the precovered surfaces (+49/53 kJ mol$^{-1}$). These differences in bonding energies can be easily explained by the following effects of the hydrogen precoverage: On the **Si-ID** surface, the molecule is thoroughly surrounded by hydrogen atoms that impose a steric repulsion, especially along the dimer rows where the distances are shorter than between the rows. This repulsion lowers the bond strength in that particular position. The **Si-H4** surface, not imposing any steric pressure along the row, has the tilting angle on both dimers reduced (see also Figure 3(a,d)) and since the Si$_{down}$ gets slightly raised in the precursor on the clean surface (comparing the height of the two visible Si$_{down}$ atoms in Figure 3(a)), the displacement needed is reduced and thus, the bonding energy gets more negative. These changes for the precursors should, however, have no effect for molecules approaching the surface from the gas phase. Adsorption into the precursor was always found to be direct and without any intermediate steps (see Figure 1 for the calculated adsorption profile on the clean 4×4 **Si** surface), so the initial sticking probability should be independent on the amount of hydrogen coverage.

A more detailed analysis of the precursor structure is found elsewhere.[30]

### 3.1.2. Reactivity Towards Covalently Bound States

Comparing the energy barriers from precursor to final state ($E^{rel}_{TS}$ values in Table 1), it can be seen that while the **Si-ID** precursor is more weakly bound than the one on the clean surface, the reaction barrier to the *on-top* structure does not change significantly through precoverage, changing merely from 9 to 8 kJ mol$^{-1}$ ($E_{ads}$) and staying constant at 10 kJ mol$^{-1}$ in Gibbs energies ($G_{ads}$). This can be understood since the reaction process is mainly dependent on the reacting dimer only, with surrounding dimers having no or only indirect influence. Since the steric repulsion by neighbouring hydrogen atoms on the precovered surface should be roughly the same in both precursor and transition state structures, their relative energies should not change much in comparison to the clean surface which is exactly what the results show.

In contrast, the barrier toward the *bridge* structure is drastically lowered through precoverage by more than 50% from 17 to 8 kJ mol$^{-1}$ ($E_{ads}$) and 17 to 7 kJ mol$^{-1}$ ($G_{ads}$), respectively. This can be explained similarly to the increased bonding energy of the precursor: During the course of the reaction, both reacting dimers have to distort and lower their tilting angle, an energy-consuming process. On the clean surface, this change is rather large (Figure 3(a,c)), while on the precovered surface, the two dimers involved are already distorted by the hydrogen atoms attached and have no need to displace as much to reach a horizontal arrangement (Figure 3(d,f)). These results also match two of our previous experimental observations:[6r] First, the measured ratio of reacted sites on the clean surface $N(bridge)/N(on\text{-}top)$ of 0.062 is very well reproduced. This ratio can be estimated by assuming thermal equilibrium in the precursor and inserting the difference in Gibbs activation energies $\Delta G_a = G_a(bridge) - G_a(on\text{-}top)$ into a Boltzmann distribution at $T$ = 300 K:

$$N(bridge)/N(on\text{-}top) = \exp(-\Delta G_a/kT) = 0.060$$

Second, the increased site-selective reactivity towards covalently bound *bridge* states at the **Si-H4** reactive sites compared to unreacted dimers is explained: Following the same arguments like above, one can calculate the ratios of relative coverage $c(bridge)/c$(Si-H4) and $c(on\text{-}top)/c$(Si-H4) to 0.018 and 0.300, respectively, from the Gibbs activation energies. This is in excellent agreement with the experimental values (0.022 and 0.286).[6r]

Reaction energies (bracketed values at $E_{final}$) are significantly more negative by precoverage, changing from −127 (**Si**) to −151 kJ mol$^{−1}$ (**Si-ID**) for the *on-top* reactions ($G_{ads}$: −119 to −143) and −110 (**Si**) to −159 kJ mol$^{−1}$ (**Si-H4**) for the *bridge* reactions ($G_{ads}$: −95 to −148). This can be explained by the fact that, on **Si-ID**, full coverage is reached and strain in the surface is released when a completely symmetric arrangement is reached. On **Si-H4** however, this is due to the lowered deformation energy of the surface mentioned before: For *bridge* adsorption on two clean dimers, both have to distort from their equilibrium geometry to form the bonds to the molecule. When a second molecule approaches this site, the distortion needed for bond formation has already been carried out, so this destabilizing component vanishes from the reaction energy.

Differences between electronic and Gibbs energies of the final states ($G_{ads} - E_{ads}$: 8-15 kJ mol$^{−1}$) are more pronounced than those for the activation energies ($G_{TS} - E_{TS}$: 1-2 kJ mol$^{−1}$). This can mainly be attributed to the change in zero-point vibrational energies (ZPVE): While the precursor and transition state structures feature low-frequency hindered translations and rotations, these motions are converted into fully vibrational modes in the final states with a higher frequency, raising the ZPVE in the final state.

Structurally, the transition states of the *on-top* reactions (Figure 2(b,e)) show the C-C bond being slightly elongated in both systems, but to the same degree. $d(C_1\text{-Si}_{down})$ is also similar at a value of 2.105 (**Si**) and 2.118 Å (**Si-ID**), but on **Si-ID**, the $C_2$ atom is still closer to this silicon atom (2.332 Å) compared to the reaction on the clean **Si** surface (2.416 Å). Final state structures for these reactions (Figure 2(c,f)) show very similar bond lengths and angles. The slight asymmetry in the C-Si bond lengths on the clean surface can be explained by the asymmetry of the adjacent tilted dimer compared to the completely symmetric hydrogen saturated dimer on the **Si-ID** reactive site. The average C-Si value on the clean **Si** surface, however, perfectly fits the value for the **Si-ID** surface (1.943 Å).

In the *bridge* reaction, the transition state structures (Figure 3(b,e)) show exactly the same C-C bond length, which is very close to the value in the precursor. This highlights that in this reaction, the transition state occurs before this bond is further weakened. One distinct difference between the transition state structures of the clean **Si** surface and the **Si-H4** one is that in the latter case, it appears way closer to the precursor structure, as can be seen by comparing the C-Si bond lengths (2.146/2.590 Å) to those in the clean surface reaction (2.168/2.840 Å). Especially the $C_2$ atom is much further away from the $Si_{down}$ in the latter case. This implies that the maximum in energy is reached way earlier along the reaction coordinate, which also coincides with the lowered energy barrier (see Table 1). Final state structures are again essentially the same, although on **Si-H4** not as symmetric as on **Si-ID**, since the molecule arranges in a *gauche* conformation for both *bridge* reactions, making the two carbon atoms geometrically inequivalent.

### 3.1.3. Comparison with Literature Data

In order to estimate the reliability of our calculated values, comparison to previous experimental and theoretical work is presented. Also, since no previous theoretical calculations investigating the reactivity used dispersion correction, it is of great interest to quantify the influence of this on the reactivity. So for better comparison, we present our values with and without dispersion correction. Furthermore, all theory values are given as electronic energies only. All gathered data are found in Table 2 (*on-top* reaction) and Table 3 (*bridge* reaction).

The most commonly reported value is $E_{ads}$ of the *on-top* final state, the theoretical results without dispersion correction range from −180 to −203 kJ mol$^{−1}$. Our corresponding value of −183 kJ mol$^{−1}$ fits very well to this. The precursor bonding energy has been calculated three times[7h,7j,7s] to be −45, −47 and −46 kJ mol$^{−1}$, respectively, and again, our value of −46 kJ mol$^{−1}$ fits excellently. Since most calculations have been done in a 2×2 cell, it can be assumed that this cell size is sufficient for both precursor and *on-top* structural motifs. The study by Cho and co-workers[7h] shows a different precursor for the *bridge* reaction with the molecular axis oriented perpendicular to the dimer bond and 2 kJ mol$^{−1}$ higher in energy (see Table 3), but as we have already shown in the calculated rotation profile (Figure 4), this second minimum is very shallow and should have a considerably shorter lifetime than the 37° orientation from which the reaction can also take place. Calculated energy barriers to the *on-top* state vary from 2 to 7 kJ mol$^{-1}$, emphasizing that small errors can have a huge influence here and that dispersion correction, although amounting for only 2 kJ mol$^{-1}$ in our calculations, makes a large relative contribution. Differences between the functionals PBE and PW91 are, as expected, only marginal, comparison between our PBE (without dispersion) values with the most recent PW91 study[7s] show exactly the same values aside from a 2 kJ mol$^{-1}$ change in the final state energy. The study by Kim and co-workers[7v] emphasized on the effect of different dispersion correction schemes on the bonding energy in these systems and since the vdW-SCS (self-consistently screened) correction, which is often considered superior to semiempirical DFT-D corrections, yields a similar adsorption energy for the final *on-top* state compared to our PBE-D3 value (−210 vs. −201 kJ mol$^{-1}$), it appears that both methods describe this aspect similarly well. The inclusion of dispersion correction can sometimes make a significant difference, in our calculations in the bonding energy of the precursor state $E_{prec}$, where the value gets lowered by almost 30 kJ mol$^{-1}$ from −46 to −74 kJ mol$^{-1}$, increasing its absolute value by 65%.

**Table 2.** Comparison of our obtained values for the *on-top* reaction on clean Si(001) with periodic DFT calculations from literature and with experiments.[a]

| Ref. | Method(size) | $E_{prec}$ | $E_{TS}$ | $E_{final}$ | $E^{rel}_{TS}$ | $E^{rel}_{final}$ |
|---|---|---|---|---|---|---|
| [7e] | PBE(2×2) | | | −186 | | |
| [7f] | PBE(2×2) | | | −203 | 5 | |
| [7h] | PBE(2×2) | −45 | −43 | −187 | 2 | −142 |
| [7j] | PW91(2×2) | −47 | −41 | −180 | 6 | −133 |
| [7q] | PW91(2×2) | | | −199 | | |
| [7s] | PW91(2×2) | −46 | −39 | −185 | 7 | −139 |
| [7v] | PBE(4×4) | | | −192 | | |
| | PBE+vdW-SCS(4×4) | | | −210 | | |
| This study | PBE(4×2) | −46 | −39 | −183 | 7 | −137 |
| | PBE-D3(4×2) | −74 | −66 | −201 | 9 | −127 |
| [6b] | Experiment | | −12 | −159 | | |
| [6q] | Experiment | | | | 12 | |
| [6t] | Experiment | | −14-19 | | | |

[a] All values in kJ mol$^{−1}$. Computational results listing electronic energies $E_{ads}$.

**Table 3.** Comparison of our obtained values for the *bridge* reaction on clean Si(001) with literature.[a]

| Ref. | Method(size) | $E_{prec}$ | $E_{TS}$ | $E_{final}$ | $E^{rel}_{TS}$ | $E^{rel}_{final}$ |
|---|---|---|---|---|---|---|
| [7f] | PBE(2×2) | | | −188 | | |
| [7h] | PBE(2×2) | −43 | −32 | −176 | 11 | −133 |
| [7q] | PW91(2×2) | | | −187 | | |
| [7s] | PW91(2×2) | | | −173 | 16 | |
| This study | PBE(4×2) | | | −173 | | |
| | PBE(4×4) | −45 | −30 | −162 | 15 | −117 |
| | PBE-D3(4×2) | | | −193 | | |
| | PBE-D3(4×4) | −74 | −57 | −184 | 17 | −110 |

[a] All values in kJ mol$^{-1}$. Computational results listing electronic energies $E_{ads}$.

The *bridge* state and the reactivity towards it (Table 3) are less well documented than the *on-top* reaction. Bonding energies of the final state vary from −173 to −188 kJ mol$^{-1}$, while our value without dispersion yields only −162 kJ mol$^{-1}$. This can, however, be explained by the smaller cell size of 2×2 in all literature calculations compared to 4×4 in ours. To verify this assumption, we also determined the value in a 4×2 cell, which has the same dimensions along the *b* axis of the cell as the calculations in the literature, and the resulting value of −173 kJ mol$^{-1}$ fits perfectly well. This emphasizes that a 4×2 or 2×2 cell is too small for this bonding motif and that non-physical interactions with images in neighbouring cells become significant and artificially lower the bonding energy by about 10 kJ mol$^{-1}$. This effect can be expected to occur for any reaction with two dimers along a dimer row, since in a 4×2 or 2×2 cell, these are the only two dimers present in the row in this system.

The reactivity in this system has also been investigated using finite cluster approaches for the surface.[7g,7i,7k,7l,7m] However, most of these calculations yield a diradical mechanism with a transition state energy above the reference zero point. Since all experiments[6b,6q,6t] and periodic calculations report this energy to be negative, the cluster approach is probably not well-suited for the questions investigated here.

Comparing the *on-top* reaction with experiments (Table 2), only one value for $E_{final}$ is available, −159 kJ mol$^{-1}$.[6b] Although our Gibbs energy value of −139 kJ mol$^{-1}$ is smaller, the deviation is in an acceptable range. Experimentally determined energy barrier fits well to our Gibbs energy value of 10 kJ mol$^{-1}$.[6q] The agreement in the transition state energy $E_{TS}$ is very good for the Gibbs energy (Table 1), while the electronic energy value is far too high. This emphasizes again the importance of thermodynamic corrections to adsorption energies.

To summarize the literature review: First, while the 2×2 cell size is appropriate for the description of the *on-top* state, it is not large enough for the *bridge* state, since the periodic boundary conditions create a geometric arrangement that artificially lowers the bonding energy. Second, dispersion correction stabilizes the adsorption modes significantly and since these forces are not described by conventional DFT functionals, they should be included in calculations of organic/semiconductor systems. The reaction barrier, too, is dependent on dispersion correction and in combination with thermodynamic corrections, a good agreement with the latest experimental values can be reached. In general, direct comparison between electronic energy differences and experimental values should be taken with care, as we have shown that dispersion- and Gibbs-corrected values are in much better agreement with experimental findings.

## 3.2. Experimental Results

Our calculations nicely confirm the previous experimental results for ethylene adsorption on the clean Si(001) surface and at low hydrogen precoverage.[6r] Here, we additionally study the site-selective reactivity of isolated dimers at high hydrogen precoverage by means of STM. The isolated dimers can be prepared by subsequent thermal annealing of a monohydride Si(001) surface. After thermal desorption of a few H$_2$ molecules and diffusion of atomic hydrogen on the surface,[27] one then yields a monohydride surface with isolated dimers (majority) and isolated dangling bonds (minority) as shown in Figure 5(a). Subsequent exposure to ethylene then leads to a distinct reduction of the number of isolated dimers while the number of single dangling bonds stays constant. Please note that the STM tip is removed during the experiment and the evaluation is not done for the very same surface area but by statistical analysis of the observed configurations. The decrease of isolated dimers indicates the passivation of their two dangling bonds by adsorbed ethylene molecules. Within the experimental errors, the expected first-order adsorption kinetics represented by the solid line in Figure 5(b) reproduces the experimental data. From the initial slope of the curve, the sticking coefficient of the isolated dimers can be estimated to be close to unity. Thus, the site-selective reactivity of the isolated dimers is very similar to the clean dimers, which is again in agreement with our calculations.

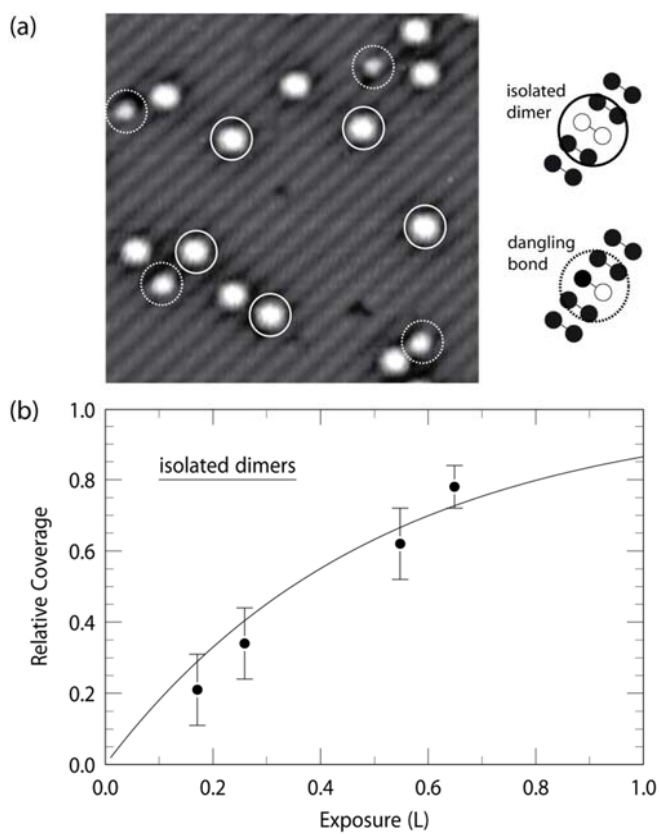

**Figure 5.** (a) STM image (-1.9 V, 0.5 nA, 15x15 nm²) of a monohydride Si(001) surface after thermal desorption of some hydrogen molecules. Bright configurations symmetric (some labelled by solid circles) and asymmetric (labelled by dashed circles) to the dimer rows are identified as isolated dimers and single dangling bonds, respectively. Subsequent exposure to ethylene leads to a decrease of the number of isolated dimers due to an *on-top* reaction while the number of single dangling bonds is constant. (b) Relative coverage of the isolated dimers with respect to the exposure of ethylene. The solid line shows the relation for a first-order adsorption kinetics.

At this point, we would like to further compare our results with the observation that the reactivity of ethylene on ethylene precovered Si(001) is strongly reduced at 0.5 ML coverage and above, when at least every second dimer is reacted by an ethylene molecule.[6c,6t,7u] With respect to the surface electronic configuration, one does not expect a major difference for the dimers being reacted either by two hydrogen atoms or a covalently bond ethylene molecule. As both our calculations and experiments do not show a decrease in reactivity for the isolated dimers between hydrogen saturated dimers, the reduced reactivity on the ethylene precovered surface is thus unlikely to result from electronic effects and has to be attributed to steric effects by the ethylene molecules on the neighbored dimers. This is confirmed by DFT calculations in a unit cell similar to **Si-ID**, but with *on-top* bound ethylene molecules instead of hydrogen atoms occupying three of the four dimers (Supplementary Information, Figure S1): The resulting precursor bonding energies, −52 kJ mol$^{-1}$ ($E_{prec}$) and +2 kJ mol$^{-1}$ ($G_{prec}$), show that at room temperature, the stabilization due to bond formation is not able to compensate for the loss in molecular entropy anymore and the state becomes thermodynamically unstable. The large difference compared to the **Si** precursor, +22 ($E_{prec}$) and +24 kJ mol$^{-1}$ ($G_{prec}$), can be attributed to steric repulsion in two different ways: First, adsorbed molecules present in the same row have to deform to make space for the impinging molecule (Figure S1). Second, even in this deformed structure, Pauli repulsion as quantified with our recently developed pEDA method is increased by 30 kJ mol$^{-1}$ compared to the clean surface while orbital interaction is only marginally weakened (+5 kJ mol$^{-1}$) and a slight stabilization in electrostatics (−13 kJ mol$^{-1}$) and dispersion (−6 kJ mol$^{-1}$) is not able to compensate for the increased Pauli repulsion (Table S4). See also the Supplementary Information for more data.

## 4. Conclusions

In summary, our calculations and experiments complement each other well by showing that isolated dimers in an environment of hydrogen-covered surface atoms show no significant change in reactivity while the reactive site on the **Si-H4** surface lowers the energy barrier tremendously and enhances the reactivity in accordance to our previous experiments. Additionally, the calculations show that the precursor bonding energies shift slightly due to hydrogen precoverage and that reaction energies are enlarged as a result of the complete saturation of adjacent dimers. Gibbs energies of bonding are presented as well and show that while temperature and pressure effects do not have a large influence on energy barriers and reaction energies, the bonding energies of adsorbed states are changed significantly, emphasizing that the thermodynamic corrections are most certainly needed to appropriately describe these quantities. In the most pronounced case of a dimer fully enclosed by molecular coverage, the precursor becomes thermodynamically unstable at room temperature. The reason for this difference to the clean surface can clearly be attributed to the Pauli repulsion between the molecules, as our bonding analysis results show.

## Acknowledgements

The authors thank Prof. U. Höfer for most fruitful discussions. This work was supported by the Deutsche Forschungsgemeinschaft (DFG) within SFB 1083. Computational resources were provided by Hochschulrechenzentrum Marburg, the HLR Stuttgart and the LOEWE-CSC Frankfurt.

**Keywords:** Si(001) surface • Ethylene • Density functional theory • Gibbs energy of adsorption • Scanning tunnelling microscopy

# ARTICLE

**Entry for the Table of Contents** (Please choose one layout)

Layout 2:

# ARTICLE

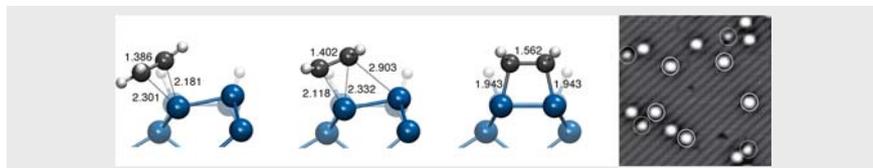

**Ethylene's favorite travel destination**: Density functional theory calculations including dispersion and thermodynamic corrections combined with bonding analysis yield an explanation for the site-specific reactivity of ethylene at distorted dangling bond configurations on Si(001). Scanning tunnelling microscopy results motivate and confirm the theoretical investigations.

*Josua Pecher, Gerson Mette, Michael Dürr\* and Ralf Tonner\**

*Page No. – Page No.*

**Site-specific reactivity of ethylene at distorted dangling bond configurations on Si(001)**